\documentclass[12pt]{iopart}
\usepackage{graphicx}
\usepackage{epsfig}
\begin{document}
\bigskip
\centerline{\bf The breakdown of the shear modulus at the glass
transition}
\bigskip
\bigskip
\centerline{U. Buchenau}
\bigskip
\centerline{\it Institut f\"ur Festk\"orperforschung,
Forschungszentrum J\"ulich} \centerline{\it Postfach 1913, D-52425
J\"ulich, Germany.}
\bigskip
\baselineskip=8truemm
\begin{quotation}
\centerline {ABSTRACT}
\par
The glass transition is described in terms of thermally activated
local structural rearrangements, the secondary relaxations of the
glass phase. The interaction between these secondary relaxations
leads to a much faster and much more dramatic breakdown of the
shear modulus than without interaction, thus creating the
impression of a separate primary process which in reality does not
exist. The model gives a new view on the fragility and the
stretching, two puzzling features of the glass transition.
\end{quotation}

The mode coupling theory of the glass transition (G\"otze and
Sj\"ogren 1992) postulates a crossover in the flow mechanism at
the critical temperature $T_c$ to thermally activated hopping on
the low-temperature side. This postulate has received strong
support by recent simulations, which showed that the system passes
from the saddle points of the energy landscape to the minima at
exactly this critical temperature (for a review see Buchenau
2003). Thus one indeed expects thermally activated flow between
$T_c$ and the calorimetric glass transition temperature $T_g$,
where the system freezes.

However, if one looks at the temperature dependence of the flow
process below $T_c$, it is dramatically faster than that of a
thermally activated process, particularly for the so-called
fragile glass formers (B\"ohmer et al 1993). This fragility seems
to be linked to a strong stretching of the shear stress
relaxation, extending over several decades in time. Both phenomena
have not yet found a generally accepted explanation.

The present paper proposes to describe the glass transition in
terms of the energy landscape concept (Goldstein 1968; Johari and
Goldstein 1970, 1971; Stillinger 1995), taking the interaction
between different thermally activated jumps into account. There is
increasing evidence (Richert 2002)  for a heterogeneous energy
landscape dynamics. In the glass phase, the energy landscape idea
has been successfully exploited in three phenomenological models,
the tunneling model (Phillips 1981) for the glass anomalies below
1 K, the soft-potential model (Parshin 1994) for the crossover to
higher temperatures and, finally, the ADWP
(Asymmetric-Double-Well-Potential, Pollak and Pike 1972) or
Gilroy-Phillips model (Gilroy and Phillips 1981) for the
description of the classical relaxation over a wide range of
barrier heights, up to barrier heights which become relevant for
the flow process, i.e. barriers of the order of 35 $k_BT_g$.

As is well known, two relaxation centers at different positions
interact via their elastic dipole moments. Here, it is proposed to
treat this interaction in terms of the weakening of the shear
modulus by the relaxation, a mean-field approach.

We describe the dynamical shear response of the glass or
undercooled liquid in terms of the {\it apparent} barrier density
function $f(V)$ defined by
\begin{equation}\label{f}
\frac{\delta G}{G\delta V}\equiv f(V),
\end{equation}
where $\delta G$ is the reduction of the shear modulus from all
relaxations with barrier heights between $V$ and $V+\delta V$ and
$G$ is the infinite frequency shear modulus. The relaxation time
$\tau_V$ of these relaxation centers is given by the Arrhenius
relation
\begin{equation}\label{arrh}
\tau_V=\tau_0\exp(V/k_BT)
\end{equation}
with $\tau_0\approx 10^{-13}s$.

One next needs the relation between the true barrier density
$f_0(V)$ and the apparent density $f(V)$. Consider a constant
shear strain $\epsilon_0$ applied at time $0$. The stress $\sigma$
after the time $t$ is given by
\begin{equation}\label{gt}
\sigma(t)=G(t)\epsilon_0=G\epsilon_0\int_0^{V_M}
\exp(-t/\tau_V)f(V)dV.
\end{equation}
Here $V_M$ is the Maxwell barrier corresponding to the terminal
relaxation time $\tau_M$ of the shear modulus, the Maxwell time.

A given relaxation center will tend to jump at its relaxation time
$\tau_V$. At the time $\tau_V$, the stress is diminished by the
factor $G_V/G$, where $G_V=G(\tau_V)$. The local distortion at the
center is assumed to be unchanged. This implies an unchanged
contribution of the center to the free energy. Thus the reduction
of the stress energy by the center is unchanged. But we reduce a
smaller stress energy, so $f(V)$ must be larger than $f_0(V)$ by
the square of the factor $G/G_V$. In a physical picture, the
enhancement is due to induced jumps of lower-barrier entities in
the neighbourhood, which reduce the resistance of the viscoelastic
medium surrounding the center.

$G_V$ is given by
\begin{equation}\label{gv}
G_V=G\int_0^{V_M} \exp(-\tau_V/\tau_v)f(v)dv\approx
G\int_V^{V_M}f(v)dv,
\end{equation}
because the double-exponential cutoff is practically a step
function. Therefore it is reasonable to define the true barrier
density function by
\begin{equation}\label{f0}
f_0(V)\equiv f(V)\left[\int_V^{V_M}f(v)dv\right]^2,
\end{equation}
since the integral on the right hand side is close to $G_V/G$.

Solving this integral equation for $f(V)$ one finds
\begin{equation}\label{fin}
f(V)=\frac{f_0(V)}{\left[3\int_V^{V_M}f_0(v)dv\right]^{2/3}}.
\end{equation}

This solution has two important properties. The first is the
1/3-rule:
\begin{equation}\label{third}
\int_0^{V_M}f_0(V)dV=\frac{1}{3}.
\end{equation}
The rigidity vanishes when the integral over the noninteracting
relaxation entities is 1/3, i.e. when the noninteracting secondary
processes would have reduced the initial shear modulus at time
zero to 2/3 of its value. The 1/3-rule has been derived
independently from energy considerations (Buchenau 2003).

The second important property is that the breakdown occurs in a
rather dramatic way, because the relaxing entities at the critical
Maxwell barrier value receive a strong enhancement, to such an
extent that one is tempted to assume a separate $\alpha$-process
which has nothing to do with the secondary relaxations. In fact,
this more or less unconscious assumption underlies most of the
present attempts to understand the glass transition (Ediger,
Angell and Nagel 1996). The above treatment shows such an
assumption to be unnecessary; what one sees at the glass
transition are simple Arrhenius relaxations of no particularly
large number density, blown up to impressive size by the small
denominator of eq. (\ref{fin}).

One can calculate the barrier density function from dynamical
mechanical measurements in the glass phase using the approximation
(Buchenau 2001)
\begin{equation}
f(V,T)\approx\frac{2}{\pi}\frac{G''}{G k_BT}, \label{fvexp}
\end{equation}
where $V$ is calculated from the condition $\omega\tau_V=1$.

\begin{figure}[b]
\hspace{-0cm} \vspace{0cm} \epsfig{file=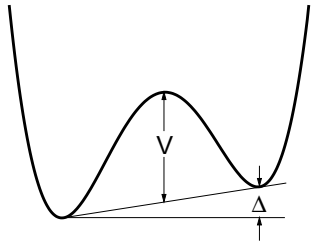,width=5
cm,angle=-90} \vspace{0cm} \caption{Asymmetric double-well
potential.}
\end{figure}

Though the potential parameters freeze at $T_g$, $f_0(V,T)$ will
still depend on temperature, even in the glass phase. This
temperature dependence can be derived from a consideration on the
free energy of an asymmetric double-well potential (Fig. 1). In
the simplest possible approximation, the free energy of the
relaxing entity is (Buchenau 2001)
\begin{equation}
F =-k_BT\ln\left[2\cosh\left(\frac{\Delta}{2k_BT} \right)\right].
\end{equation}
The Boltzmann factor $\exp(-F/k_BT)$ at $T_g$ supplies the
asymmetry distribution
\begin{equation}\label{boltz}
P(\Delta)\sim\cosh(\Delta/2k_BT_g),
\end{equation}
so the number density of relaxing entities does indeed {\it
increase} with increasing $\Delta$; it is not really constant. If
one integrates over the asymmetries at a lower temperature, one
finds a diminution of the effective $f_0(V,T)$ by a factor of up
to $\pi/2$ in the low temperature limit. A useful approximation
below $T_g$ is
\begin{equation}\label{tfac}
\frac{f(V,T_g)}{f(V,T)}= \left[2{\rm arctan\ (e}^{
V/2k_BT})-\frac{\pi}{2}\right]{\rm e}^{-(T_g/T)^2{\rm ln}(\pi/2)}
\end{equation}
This factor has to be taken into account to calculate $f(V,T_g)$
from measurements of the mechanical loss in the glass phase. Once
$f(V)$ is known, $f_0(V)$ can be obtained from eq. (\ref{f0}).

\begin{figure}[b]
\hspace{-0cm} \vspace{0cm} \epsfig{file=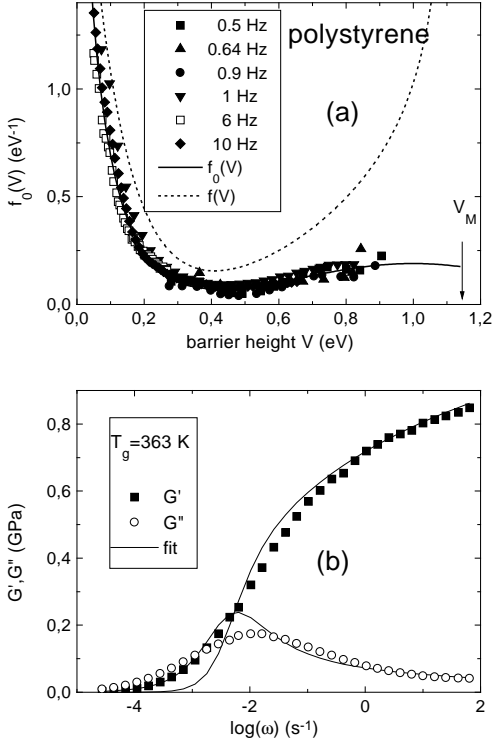,width=8
cm,angle=0} \caption{ (a) $f_0(V,T_g)$ for polystyrene from
torsion pendulum data, together with the corresponding $f(V,T_g)$
(b) measured and calculated dynamical shear data at the glass
transition (references see text).}
\end{figure}

Fig. 2 (a) shows $f_0(V,T_g)$ for polystyrene determined in this
way from six torsion pendulum data at 0.5, 0.64, 0.9, 1, 6 and 10
Hz (Illers and Jenckel 1958; Schwarzl 1990a; Schmieder and Wolf
1953; Schwarzl 1990b; Sinnott 1959; Hartwig and Schwarz 1968). The
data were fitted by the six-parameter expression
\begin{equation}\label{f0fit}
f_0(V,T_g)=\frac{2C}{W^{3/4}V^{1/4}}{\rm e}^{-V/V_0}+a_1{\rm
e}^{-g1(V-p1)^2}.
\end{equation}
Two of the parameters, the energy $W$ and the small number $C$,
can be taken from soft-potential fits of the low-temperature
anomalies (Ramos and Buchenau 1997), because the first term is the
barrier density of the soft-potential model (Parshin 1994), with
an exponential cutoff of the soft-potential expectation at a
limiting barrier $V_0$. The second term is just an appropriate
gaussian with amplitude $a_1$ and width parameter $g_1$ centered
at the position $p_1$. The results of the fit are compiled in
Table I.

\bigskip

\ \ \ \ \ \ Table I: Fit parameters for $f_0(V,T_g)$.

\

\begin{tabular}{|c|c|c|c|c|c|c|}
\hline
  substance& $C$ & $W$ & $V_0$ & $a_1$ & $g_1$ & $p_1$ \\ \hline
         &$10^{-4}$  & $meV$ & $eV$ & $eV^{-1}$ & $eV^{-2}$ & $eV$ \\
         \hline
  $a-SiO_2$ & 2.65 & 0.34 & 0.04 & 0.76 & 0.54 & 5.45 \\ \hline
  polystyrene& 7.1 & 0.159 & 0.1 & 0.19 & 4.0 & 1.0 \\ \hline
\end{tabular}

\bigskip

With these parameters, the integral over $f_0(V,T_g)$ extrapolates
to 1/3 at $V_M=1.1 eV$, consistent with the experimental $T_g$ of
373 K, defining $T_g$ as the temperature where the terminal
relaxation time $\tau_t$ reaches 100 s. This is the first time a
glass temperature could be calculated from glass data alone (it is
true that the glass temperature enters already in the calculation,
but one can iterate, and the procedure converges quickly).

Even more impressive, the same parameter combination provides a
very reasonable fit of the measured (Donth \etal 1996)
$G'(\omega)$ and $G''(\omega)$ at the glass temperature (for this
sample at 363 K). This is shown in Fig. 2 (b), calculated with
$G=1.55$ GPa, the value obtained from the Brillouin transverse
sound velocity (Strube, private communication) of 1071 m/s, taking
the weakening by $f_0(V)$ at the Brillouin frequency into account.
The sharp barrier cutoff at $V_M$ of the model seems to be too
sharp for this polymer case, possibly because of the crossover
from enthalpic to entropic restoring forces.

\begin{figure}[b]
\hspace{-0cm} \vspace{0cm} \epsfig{file=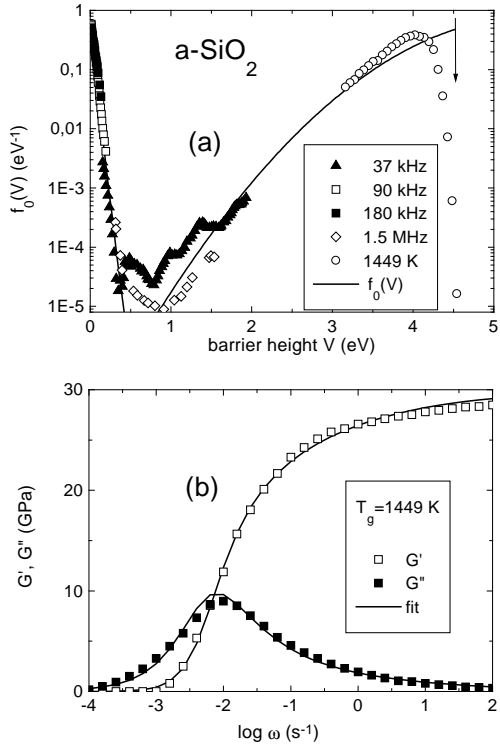,width=8
cm,angle=0} \vspace{0cm} \caption{ (a) $f_0(V,T_g)$ for $SiO_2$
from measurements at four frequencies and from the dynamical glass
transition data (b) measured and calculated dynamical shear data
at the glass transition (references see text).}

\end{figure}

Fig. 3 (a) shows the barrier density function in the strong glass
former $SiO_2$. From measurements at 90 and 180 kHz (Cahill and
van Cleve 1989; Keil, Kasper and Hunklinger 1993), one finds again
the soft-potential model expectation
$f_0(V,T_g)=2C/W^{3/4}V^{1/4}$ at very low barriers, with $C$ and
$W$ taken from fits to the low-temperature thermal conductivity
and specific heat (Ramos and Buchenau 1997). There is a pronounced
exponential cutoff over more than four decades leading to very low
mechanical damping at and above room temperature, the lowest
observed in any glass so far. Data close to $T_g$ are rather
scarce. Therefore it turned out to be necessary to include the
data (Mills 1974) for $G'(\omega)$ and $G''(\omega)$ at $T_g$ into
the fit for the high-barrier gaussian in eq. (\ref{f0fit}). Thus,
one cannot predict $T_g$ from glass phase data alone as in
polystyrene. The fulfillment of the 1/3-rule required $G=30.2$ GPa
for the fit of Fig. 3 (b), while transverse wave Brillouin data
(Bucaro and Dardy 1974) suggest a value of $33.7$ GPa.
Nevertheless, having fitted $f_0(V)$ both at the low- and
high-barrier end, it is gratifying to see the good agreement of
the very low damping at and above room temperature for 37 kHz
(Marx and Sivertsen 1953) and 1.5 MHz (Fraser 1970) in Fig. 3 (a),
because these data were not used to obtain the fit. Also, Fig. 3
(b) shows that in this case the sharp cutoff gives a much better
fit of the stretching than in polystyrene.

Note that not all glass formers can be fitted by the simple form
of eq. (\ref{f0fit}); many glass formers show several separate
secondary relaxation peaks.

Finally, let us turn to the fit of the fragility. This requires
the temperature dependence of a whole function, namely $f_0(V,T)$
above $T_g$. Here, the estimate for this dependence is based on
the Boltzmann factor of eq. (\ref{boltz}). Obviously, the rise of
the Boltzmann factor with increasing asymmetry $\Delta$ cannot go
on forever. We assume that it stops at $\Delta=b_oV$, where $b_o$
is a fit parameter of order 1. One expects a return to a kind of
low-energy glass ground state when the asymmetry gets as large as
the barrier height.

\begin{figure}[b]
\hspace{-0cm} \vspace{0cm} \epsfig{file=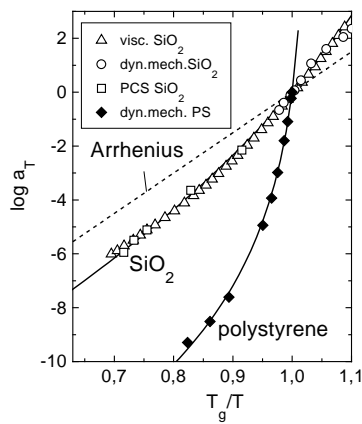,width=5
cm,angle=0} \vspace{0cm} \caption{Angell plot for polystyrene and
$SiO_2$. The dashed line shows the Arrhenius law, the continuous
lines the model fits explained in the text.}
\end{figure}

This assumption allows to calculate a partition function for a
given barrier height. From this partition function, $f_0(V,T)$
should follow the equation
\begin{equation}\label{f0rise}
\frac{f_0(V,T)}{f_0(V,T_g)}=\frac{T_g}{T}\frac{{\rm
sinh}(b_oV/2k_BT_g)}{{\rm sinh}(b_oV/2k_BT)}
\end{equation}
for temperatures above $T_g$.

With this equation, one can calculate the shift factor
$a_T=\tau_M(T)/\tau_M(T_g)$ from the 1/3-rule, eq. (\ref{third}).
The result is compared in Fig. 4 to measurements (Mills 1974;
Bucaro and Dardy 1977; Plazek and O'Rourke 1971). The fit
parameter $b_o$ was 1.4 for polystyrene and 0.33 for silica. The
difference in fragility does not only stem from this difference,
but also from the large difference in $f_0(V_M)$ at $T_g$: a rise
in the number of relaxing entities shifts the Maxwell barrier much
farther in polystyrene than in silica.

This fragility fit may be too simple-minded, but it points the way
to a new understanding. If $f_0(V,T)$ increases with increasing
temperature, the Maxwell barrier will decrease. Once one accepts
the idea of a relatively high free energy of a symmetric
double-well potential in the undercooled liquid, the fragility
ceases to be a puzzling property.

\section*{References}
\begin{harvard}
\item[] B\"ohmer R., Ngai K. L., Angell C. A. and
Plazek D. J. 1993 {\it J. Chem. Phys.} {\bf 99} 4201
\item[] Bucaro J. A. and Dardy H. D. 1974 {\it J. Appl. Phys.} {\bf 45}, 5324
\item[] Bucaro J. A. and Dardy H. D. 1977 {\it J. Non-Crystalline
Solids} {\bf 24}, 121
\item[] Buchenau U. 2001 {\it Phys. Rev. B} {\bf 63}, 104203
\item[] Buchenau U. 2003 preprint cond-mat/0209172,
{\it Energy landscape - a key concept in the dynamics of liquids
and glasses}; {\it J. Phys.: Condens. Matter} {\bf 15}, S955
\item[] Cahill D. G. and van Cleve J. E. 1989 {\it Rev. Sci. Instrum.} {\bf
60}, 2706
\item[] Donth E., Beiner M., Reissig S., Korus J.,
Garwe F., Vieweg S., Kahle S., Hempel E. and Schr\"oter K. 1996
{\it Macromolecules} {\bf 29}, 6589-6600
\item[] Fraser D. B. 1970, {\it J. Appl. Phys.} {\bf 41}, 6
\item[] Geis N., Kasper G. and Hunklinger S. 1986, in {\it Non-metallic
Materials and Composites at Low Temperatures 3}, ed. by Hartwig G.
and Evans D., p. 99 (Plenum Press, New York)
\item[] Gilroy K. S. and Phillips W. A. 1981 {\it Phil. Mag. B} {\bf 43}, 735
\item[] G\"otze W. and Sj\"ogren L. 1992 {\it Rep. Prog. Phys.} {\bf
55} 241
\item[] Goldstein M. 1968 {\it J. Chem. Phys.} {\bf 51}, 3728
\item[]Hartwig G. and Schwarz G. 1986, in {\it Nonmetallic Materials and
Composites at Low Temperatures 3}, ed. by Hartwig G. and Evans D.,
p. 117 (Plenum Press, New York)
\item[] Illers K. H. and Jenckel E. 1958 {\it Rheol. Acta} {\bf
1}, 322
\item[] Keil R., Kasper G. and Hunklinger S. 1993 {it J. Non-Cryst. Sol.} {\bf 164-166}, 1183
\item[] Johari G. P. and Goldstein M. 1970 {\it J. Chem. Phys.} {\bf 53} 2372
\item[] Johari G. P. and Goldstein M. 1971 {\it J. Chem. Phys.} {\bf 55} 4245
\item[] Marx J. W. and Sivertsen J. M. 1953 {\it J. Appl. Phys.} {\bf 24}, 81
\item[] Mills J. J. 1974 {\it J. Noncryst. Solids} {\bf 14}, 255
\item[] Parshin, D. A. 1994 {\it Phys. Solid State} {\bf 36}, 991
\item[] Phillips, W. A. (ed.) 1981, {\it Amorphous Solids: Low temperature
properties}, (Springer, Berlin)
\item[] Plazek D. J. and O'Rourke V. M. 1974 {\it J. Polym. Sci:
Part A-2} {\bf 9}, 209
\item[] Pollak M. and Pike G. E. 1972 {\it Phys. Rev. Lett.} {\bf
28} 1449
\item[] Ramos, M. A., and Buchenau, U. 1997 {\it Phys. Rev. B} {\bf 55}, 5749
\item[] Richert R. 2002 {\it J. Phys.: Condens. Matter} {\bf 14}
R703-R738
\item[] Schmieder K. and Wolf K. 1953 {\it Kolloid-Z.} {\bf 136},
157
\item[] Schwarzl F. R. 1990a {\it Polymermechanik} (Springer, New York), Fig.
5.16
\item[] Schwarzl F. R. 1990b {\it Polymermechanik} (Springer, New York),
Fig. 6.12
\item[] Sinnott K. M. 1959 {\it J. Polym. Sci.} {\bf 128}, 273
\end{harvard}

\end{document}